\documentstyle[12pt,fleqn]{article}
\def\thesection{}
\def\be{\begin{equation}}
\def\ee{\end{equation}}
\def\ba{\begin{eqnarray}}
\def\ea{\end{eqnarray}}
\def\re#1{{$^{\ref{#1}}$}}

\def\la{\mathrel{\mathpalette\fun <}}
\def\ga{\mathrel{\mathpalette\fun >}}
\def\fun#1#2{\lower3.6pt\vbox{\baselineskip0pt\lineskip.9pt
        \ialign{$\mathsurround=0pt#1\hfill##\hfil$\crcr#2\crcr\sim\crcr}}}

\def\f{\phi}
\def\s{\sigma}
\def\e{\eta}
\def\l{\lambda}
\def\rf{\langle \f \rangle}

\def\a{\alpha}
\def\fp{\phi^{\prime}}

\def\ep{\varepsilon}
\def\bk{{\bf k}}
\def\br{{\bf r}}
\def\beq{\begin{equation}}
\def\eeq{\end{equation}}
\begin{document}
\begin{titlepage}
\vspace*{-62pt}
\begin{flushright}
FERMILAB-Pub-92/222-A \\
NSF-ITP-92-102\\
June 1992
\end{flushright}
\vspace{0.5in}
\centerline{\large \bf Critical behavior in the electroweak phase transition}
\vspace{.5in}
\centerline{Marcelo Gleiser}
\vspace{.1in}
\centerline{\it Department of Physics and Astronomy, Dartmouth College,
Hanover, NH~~03755}
\vspace{.3in}
\centerline{Edward W.~Kolb}
\vspace{.1in}
\centerline{\it NASA/Fermilab Astrophysics Center}
\centerline{\it Fermi National Accelerator Laboratory, Batavia, IL~~60510}
\centerline{\it and Department of Astronomy and Astrophysics, Enrico Fermi
        Institute}
\centerline{\it The University of Chicago, Chicago, IL~~ 60637}
\vspace{.5in}
\baselineskip=24pt
\centerline{\bf Abstract}
\begin{quotation}
We examine the behavior of the standard-model electroweak phase
transition in the early Universe. We argue that close to the critical
temperature it is possible to estimate the {\it effective} infrared corrections
to the 1-loop potential using well known $\varepsilon$-expansion results from
the theory of critical phenomena in 3 spatial dimensions. The theory with the
$\varepsilon$-corrected potential exhibits much larger fluctuations in the
spatial correlations of the order parameter, considerably weakening the
strength of the transition.

\end{quotation}

\vspace{0.1in}

\centerline{PACS numbers: 98.80--k, 98.80.Cq, 12.15.--y}
\centerline{e-mail: gleiser@peterpan.dartmouth.edu; rocky@fnas01.fnal.gov.}

\end{titlepage}

\baselineskip=24pt
\thesection{\bf I. INTRODUCTION}
\setcounter{section}{1}
\setcounter{equation}{0}
\vspace{24pt}

The possibility that the baryon number of the Universe can be
generated at the electroweak phase transition has triggered a lot of
interest in understanding the dynamics of weakly first-order phase
transitions in the early Universe.  Since the original work of Kuzmin,
Rubakov, and Shaposhnikov\re{KRS} a great deal of effort has been
dedicated to the construction of viable models that could generate the
required baryon asymmetry.\re{EWB}

As is well-known, one of the necessary ingredients of a successful
baryogenesis scenario is a departure from equilibrium.  Current
scenarios of electroweak baryogenesis rely on the first-order nature
of the phase transition to generate the required out-of-equilibrium
conditions in the decay of the symmetric metastable phase by the
nucleation of bubbles of the broken-symmetric phase.  Baryon number is
generated by the expansion of the bubble wall either by the scattering
of heavy fermions off the wall,\re{CKN} or by the unwinding of
topologically non-trivial configurations in its neighborhood.\re{TZ}

Although it is now generally believed that a successful baryogenesis
scenario at the electroweak scale requires a departure from the
minimal electroweak model, understanding the dynamics of the electroweak
phase transition is a crucial ingredient for any viable scenario.  One
of the main obstacles to a comprehensive study of the electroweak
phase transition is our lack of knowledge of the correct effective
potential that describes the system in the vicinity of the critical
temperature, $T_C$.  Problems due to infrared divergences have been
known since the original work of Dolan and Jackiw and
Weinberg.\re{FINITET}\ \ For the current limits on the masses of the
Higgs and top quark, the 1-loop effective potential predicts a weak
first-order transition.$^{\ref{AH}-\ref{DINE} }$ This is somewhat
unsettling, because we know that weak first-order transitions have
large infrared divergences which are not accounted for by the 1-loop
calculation. In other words, if the 1-loop potential predicts a weak
first-order transition, chances are that the actual transition is even
weaker, if not actually second order.  It is thus important to
incorporate the infrared corrections to the effective potential. In
fact, a few recent works have incorporated some infrared corrections
caused by the vanishing of vector boson masses near $\phi=0$ by
summing over ring, or daisy, diagrams.\re{BH}\ \ As clearly shown in
the paper by Dine {\it et al.,}\re{DINE} these corrections decrease
the effective tunneling barrier for decay, weakening the strength of
the transition. The validity of the ring-improved effective potential
for the temperatures of interest relies on cutting off higher-order
contributions by invoking a non-perturbative magnetic plasma mass,
$M_{\rm plasma}$, for the gauge bosons such that the loop expansion
parameter, $g^2T/M_{\rm plasma}$, is less than 1. Since this
non-perturbative contribution is not well understood at present, one
should take the results from the ring-improved potentials with some
caution.\re{ARNOLD}

In addition to infrared problems caused by the vanishing of the vector
boson masses in the symmetric phase, for sufficiently weak transitions
we point out that one must take into account infrared divergences
caused by the small Higgs mass if $M_H\ll T$.  We will review the
well-known formalism for field theory at high temperatures.  We point
out that the loop expansion parameter diverges as the Higgs mass
vanishes.  This means that diagrams contributing to the 1-loop
potential that are unimportant at either high or low temperatures may
be dominant around the critical temperature.  For instance, at zero
temperature the loop expansion parameter for the Higgs loops is
$\lambda$, the Higgs self coupling.  However for the 3-dimensional
effective field theory at high temperature, the loop expansion
parameter is $\lambda T/M_H(T)$ for the Higgs loops.  For $T\gg
M_H(T)$, the loop expansion is not under control.  We point out that
this is exactly the situation in the standard electroweak model
between the critical temperature and the spinodal temperature where
$M_H(T)$ vanishes.

In this paper, we will attempt to estimate the magnitude of the
infrared corrections to the 1-loop electroweak potential near the
critical point using a familiar technique from condensed-matter
physics. In particular, we will argue that near the critical point it
is possible to estimate the fluctuations in the spatial correlations
of the magnitude of the scalar field by using well known results from
the theory of critical phenomena. We will show that the effective
corrections to the critical exponent that controls the behavior of the
correlation length for the electroweak model can be approximated by
considering an associated Ginzburg--Landau (G--L) model just above its
critical temperature.  This approach has been successfully implemented
by De Gennes in the study of liquid crystals,\re{DEGENNES} and
recently by Fern\'andez {\it et al.} in the study of the
two-dimensional 7-states Potts model which exhibits a weak first-order
transition.\re{POTTS}\ \ In order to make our approach clear it is
instructive to examine the critical behavior of the 1-loop effective
potential for the electroweak model.

The 1-loop finite-temperature corrections to the electroweak potential
have been studied in detail in the literature, most recently by
Anderson and Hall.\re{AH}\ \ They showed that a high temperature
expansion of the 1-loop potential closely approximates the full
1--loop potential for $M_H\la 150$ GeV and $M_T\la 200$ GeV.  (It is
important to differentiate between the finite temperature Higgs mass,
$M_H(T)$ and the zero-temperature Higgs mass, $M_H$.)  They obtain for
the potential
\beq
\label{eq:VEW}
V_{{\rm EW}}(\f,T)=D\left (T^2-T_2^2 \right )\f^2-ET\f^3+{1\over 4}\l_T\f^4,
\eeq
where the constants $D$ and $E$ are given by $D=\left
[6(M_W/\s)^2+3(M_Z/\s)^2+6(M_T/\s)^2\right ]/24$, and $E=\left
[6(M_W/\s)^3+3(M_Z/\s)^3\right ]/12\pi$. Here $T_2$ is the temperature
at which the origin becomes an inflection point ({\it i.e.}, below
$T_2$ the symmetric phase is unstable and the field can classically
evolve to the asymmetric phase by the mechanism of spinodal
decomposition), and is given by
\beq
\label{eq:T2}
T_2=\sqrt{(M_H^2-8B\s^2)/4D}\ ,
\eeq
where the physical Higgs mass is given in terms of the 1-loop
corrected $\l$ as $M_H^2=\left (2\l+12B\right )
\s^2$, with $B=\left (6M_W^4+3M_Z^4-12M_T^4\right )/64\pi^2\s^4$.
We use $M_W=80.6$ GeV, $M_Z=91.2$ GeV, and $\s=246$ GeV. The
temperature-corrected Higgs self-coupling is
\beq
\l_T=\l-{1\over {16\pi^2}}\left [
\sum_Bg_B\left ({{M_B}\over {\s}}\right )^4
{\rm ln}\left (M_B^2/c_BT^2\right )-\sum_Fg_F\left ({{M_F}\over {\s}}
\right )^4{\rm ln}\left (M_F^2/c_FT^2\right )\right]
\eeq
where the sum is performed over bosons and fermions (in our case only
the top quark) with their respective degrees of freedom $g_{B(F)}$,
and ${\rm ln}c_B=5.41$ and ${\rm ln}c_F=2.64$.

Apart from $T_2$, there will be two temperatures of interest in the
study of the phase transition. For high temperatures, the system will
be in the symmetric phase with the potential exhibiting only one
minimum at $\rf=0$. As the Universe expands and cools an inflection
point will develop away from the origin at
\beq
\label{eq:FINF}
\f_{{\rm inf}}=3ET_1/2\l_T,
\eeq
where $T_1$ is given by
\beq
\label{eq:T1}
T_1=
T_2/\sqrt{1-9E^2/8\l_TD}\ .
\eeq
For $T<T_1$, the inflection point separates into a
local maximum at $\f_-$ and a local minimum at $\f_+$,
with $\f_\pm=\{ 3ET\pm
\left [9E^2T^2-8\lambda_TD(T^2-T_2^2)\right ]^{1/2}\}/2\lambda_T$.
At the critical temperature
\beq
T_C =T_2/\sqrt{1-E^2/\l_TD}\ ,
\label{eq:TC}
\eeq
the minima have the same free energy, $V_{{\rm EW}}(\f_+,T_C)=V_{{\rm
EW}}(0,T_C)$. (Note that $V(\f,T)$ is the homogeneous part of the free
energy density whose minima denote the equilibrium states of the
system. Accordingly, in this work we freely interchange between
calling $V(\f,T)$ a potential and a free energy density.)

In Fig.~1 we show the electroweak potential at temperatures
$T\gg T_1,~T_1,~T_C,$ $T_2$, and $T=0$. The difference between the
temperatures $T_1,\ T_C$, and $T_2$ is determined by the parameter
\be
x=E^2/\l_TD.
\ee
This parameter is shown in Fig.~2 for different values of $M_H$ and
$M_T$.  Clearly $x\ll 1$ for the minimal electroweak model, so we
can write the approximate relations
\ba
T_C & \simeq & T_2(1+x/2) \nonumber \\
T_1 & \simeq & T_2(1+9x/16).
\ea

It is useful to understand why the transition is first order; {\it
i.e.}, why at $T_C$ there is a barrier between the high-temperature
phase and the low-temperature phase.  It has been appreciated for a
long time that a pure $\lambda\phi^4$ theory is equivalent to a
Ginzburg--Landau theory, which has a second-order phase transition.
The reason the electroweak theory is first order, rather than second
order, is that there is an additional attractive force between scalar
particles mediated by the vector bosons.  This additional attractive
force leads to a condensate of the Higgs field at a temperature
slightly above $T_2$.\re{DINE}\ \ $T_2$ and $T_C$ would be the same (a
second-order transition) in the absence of gauge boson interactions.
(Note that as $E\rightarrow 0$, {\it i.e.}, as vector interactions
are turned off, $T_C\rightarrow T_2$.)

The whole picture of bubble nucleation relies on the behavior of
$V_{{\rm EW}} (\f,T)$ between $T_C$ and $T_2$. In the standard
picture, one assumes that the system is in a near-homogeneous state
around its equilibrium value (in this case $\langle \f \rangle=0$), so
that large thermal fluctuations in the spatial correlations of $\f$
are exponentially suppressed above the scale of the thermal
correlation length, $\xi(T)$,
\beq
\label{eq:XI}
\xi^{-2}(T)\equiv
M_H^2(T)={{\partial^2 V_{{\rm EW}}(\f=\rf ,T)}\over {\partial \f^2}}.
\eeq

In this case, for some temperature $T_C>T>T_2$, critical bubbles of
the broken-symmetric phase appear and expand.  They eventually collide
with other bubbles, converting the symmetric phase into the
broken-symmetric phase.

For the electroweak potential the difference between $T_C$ and $T_2$
is very small: $\eta_2(T_C)\equiv (T_C-T_2)/(T_C+T_2)\sim x/4\ll 1$.
The transition is predicted to be weakly first order. As mentioned
above, infrared corrections to the 1-loop potential can be very
important due to its flatness (small mass) around $\f=0$.  As we shall
see below, the loop expansion parameter for the Higgs loops at high
temperatures is not $\lambda$, but $\lambda_T T/M_H(T)$.  We can
estimate where this will become large for the standard electroweak
model.  Before starting, it is helpful to note that the
temperature-corrected Higgs self-coupling, $\lambda_T$, is
approximately equal to the {\em tree-level} Higgs self-coupling,
$\lambda_0=M_H^2/2\sigma^2$.  (It is easy to see why the
temperature-dependent logarithmic correction approximately cancels the
zero-temperature 1-loop logarithmic correction if one adopts the
renormalization scheme of Ref.\ \ref{DINE}.)  For the electroweak
potential near $T_C$, $M_H^2(T_C)=2D(T_C^2-T_2^2)$. Since
$T_2^2/T_C^2=1-E^2/\lambda_TD$, $M_H(T_C)=T_CE\sqrt{2/\lambda_T}$.
Therefore at $T_C$ the loop expansion parameter is
$\lambda_TT_C/M_H(T_C)=\lambda_T^{3/2}/E\sqrt{2}$. Now as discussed
above, to a reasonable accuracy $\lambda_T=M_H^2/2\sigma^2$ (here, of
course, $M_H$ is the zero-temperature mass).  Thus
\be
\lambda_TT_C/M_H(T_C)\simeq M_H^3/4E\sigma^3\sim 1.68 (M_H/100\ {\rm GeV})^3.
\ee
For $M_H$ greater than about 84 GeV, at $T_C$ the expansion parameter
exceeds unity.  Between $T_C$ and $T_2$ the mass goes to zero, so the
corrections are even larger.

The question we would like to address in this paper is, can we
estimate the magnitude of the infrared corrections in a simple way?
Since we are interested in the behavior of the system around $\rf=0$
for $T_C\geq T\geq T_2$, we will show that it is possible to map the
electroweak potential {\it in a small neighborhood around} $\f=0$ to
an effective Ginzburg--Landau (G--L) theory which exhibits a
second-order phase transition at $T_2$.  The critical behavior of this
model has been extensively studied in the seventies using
renormalization group (RG) techniques pioneered by Wilson.\re{WILSON}\
\ In particular, infrared corrections to the G--L model which are
important around the critical temperature have been computed using
$\varepsilon$-expansion techniques. The net result is that the
magnitude of fluctuations on the spatial correlations of the order
parameter calculated by mean-field theory (which we will show is
equivalent to the 1-loop potential) is largely underestimated. We will
obtain the corrections to the G--L model and map it back to the
electroweak potential in an attempt to estimate the infrared
corrections to the 1-loop result. We will show that the corrections to
tunneling rates can be very large, indicating the failure of the
na\"{\i}ve 1-loop potential to describe the dynamics of the
transition.

This paper is organized as follows. In Section 2 we follow
Ginsparg\re{GINSPARG} and show how we can study the finite temperature
behavior of a field theory in 4 dimensions ($d=4$) by looking at the
static (zero Matsubara frequency) mode of an effective theory in
$d=3$.  We then study the critical behavior of a G--L model in $d=3$,
emphasizing the infrared corrections to the correlation length
obtained by $\varepsilon$-expansion methods. In Section 3 we establish
the connection between the critical behavior of the electroweak
potential and an associated G--L model. We do it using two different G--L
models, showing that they give the same results. We obtain an
$\varepsilon$-corrected mass and study its effects on the nucleation
rate. In Section 4 we strengthen our arguments by estimating the
thermal dispersion of the scalar field around the origin and by
repeating our recent calculation for the nucleation rate for
non-perturbative sub-critical fluctuations. Based on the infrared
corrections obtained in Section 3, we argue that sub-critical bubbles
offer a simple estimate of the failure of the 1-loop result. We end in
Section 5 with general comments on the nature of weak first-order
transitions.

\vspace{48pt}
\thesection{\bf II. CRITICAL BEHAVIOR OF $\f^4$ FIELD THEORY}
\setcounter{section}{2}
\setcounter{equation}{0}
\vspace{18pt}

In order to study the critical behavior of a $\f^4$ scalar field theory
we follow Ginsparg\re{GINSPARG} in reducing the theory to an effective
theory of the static mode of the scalar field in $d=3$ dimensions.
The generating functional  in the presence of a source $J(x)$ for a zero

temperature scalar field theory in Euclidean ($t=-i\tau$) space-time

is (we use $\hbar=c=1$)
\beq
\label{eq:Z0}
Z[J]=\int [{\cal D}\f]\, \exp\left\{-\int d^4\!x\left[
{1\over 2}\left (\partial\f
\right )^2-{1\over 2}m^2\f^2+{{\l}\over 4}\f^4\right]-\int d^4\!xJ\f\right\}.
\eeq
In order to study the theory at finite temperature we take the Euclidean time
to be periodic in $\beta$, and sum only over periodic paths with
$\f(0,{\bf x})=\f(\tau,{\bf x})$, as is well known.\re{FINITET}\ \
Due to the periodic behavior in $\tau$ we can expand the scalar field as
\beq
\label{eq:EXP}
\f(\tau,{\bf x})={1\over {\beta}}\sum_{n=-\infty}^{+\infty}
\int {{d^3\!k}\over {(2\pi)^3}}\exp\left(i\omega_n\tau+i{\bf k\cdot x}\right)
\f_n({\bf k});
\qquad \omega_n=2\pi n/\beta.
\eeq
By rescaling the field $\f_n({\bf k})$ by $\beta^{-1/2}$, and

separating the static ($n=0$) mode from the rest, we obtain,

\ba
Z[J]&=&\int_{{\rm periodic}}[{\cal D}\f]\,
{\rm exp}\left\{ -\int_{{\bf k}}{1\over 2}
    \left ({\bk}^2-m^2\right )\f_0(\bk )\f_0(-\bk ) \right.\nonumber\\
& & \!\!\!\!\!\!\!
-\sum_{n\neq 0}\int_{\bk }{1\over 2}\left[\left (2\pi n/\beta\right)^2
+\bk^2-m^2\right] \f_n(\bk )\f_{-n}(-\bk ) +\int J\f\nonumber\\
& & \!\!\!\!\!\!\!
-\frac{\l/\beta}{4}\sum_{n,n',n''=-\infty}^{+\infty}\left.\int_{\bk ,\bk'
,\bk ''}
\f_n(\bk )\f_{n'}(\bk ')\f_{n''}(\bk '')\f_{-n-n'-n''}(-\bk -\bk '-\bk '')
\right\}
\ea
where $\int_{\bk}=\int d^3\!k/(2\pi)^3$.  The effective $d=3$ theory
is obtained by summing over all the $n\neq 0$ modes. Perturbatively,
this means that all internal lines in the Feynman diagrams will
correspond to sums over the $n\neq 0$ modes, and the external lines
are given only by the $n=0$ mode. This way the higher modes will
contribute to the mass, wave-function renormalizations, and to the
$N$-point function for the effective theory of the $n=0$ mode.  It is
then possible to construct an effective Lagrangian ${\cal L}_{{\rm
eff}}$ for the $d=3$ theory by a systematic perturbation expansion in
$\l$. The leading contribution to the 2-point function is given by the
tadpole diagram obtained by summing over the higher modes in the
Feynman propagator. One obtains to leading order,
\beq
\label{eq:LMF}
{\cal L}_{{\rm eff}}={1\over 2}\left (-m^2+{{\l}\over {4\beta^2}}+\dots
\right )\f_0(\bk )\f_0(-\bk )+{{\l/\beta}\over 4}\f_0^4+\dots .
\eeq
Note that in the effective $d=3$ theory the coupling $\l$ is
dimensionful with the temperature naturally setting its
dimensionality. This temperature dependence will not be relevant for
the discussion of the critical behavior of the theory and will be
absorbed into the definition of $\l$. The static critical behavior of
the original $d=4$ theory is completely embodied in the effective
theory for the $n=0$ mode in $d=3$.  To leading order in $\l$, this
theory exhibits a phase transition at a critical temperature
\beq
\label{eq:T}
T_C^2=1/\beta_C^2={{4m^2}\over {\l}}\left (1+O(\l)+\dots \right ).
\eeq
It turns out that the critical behavior of the theory above, known as
the G--L model, has been extensively studied using the $\ep$-expansion
to incorporate higher-order infrared effects.  From ${\cal L}_{{\rm
eff}}$ we can obtain the effective 3-dimensional potential to leading
order in $\l$,
\beq
\label{eq:VGl}
V_{{\rm G-L}}(\f,T)={{m^2(T)}\over 2}\f^2+
                  {{\l}\over 4}\f^4; \qquad m^2(T)\equiv {{\l}\over 4}
\left (T^2-T_C^2\right ),
\eeq
where $\f({\bf x})$ is the static scalar field, which is the relevant
order parameter in equilibrium.

As is well-known,\re{BREZIN} this theory exhibits a second-order phase
transition at $T_C$; above $T_C$ the left-right symmetry is exact and the
equilibrium value of $\f$ is $\rf$=0. Below $T_C$ the symmetry is broken and
the equilibrium value of $\f$ is $\rf=\pm \left[m^2(T)/\l\right]
^{1/2}$. In the thermodynamic limit, the system will eventually settle at one
value of $\f$, since any interface is energetically unfavored. Of course
$\rf$ only gives information about the homogeneous behavior of $\f$. Typically,
there will be fluctuations around $\rf$ which are correlated within the
correlation length scale defined in Eq.~(\ref{eq:XI}). For temperatures above
and below $T_C$ (denoted by $+$ and $-$ respectively) we obtain from

Eq.~(\ref{eq:VGl})
\beq
\label{eq:XI+}
\xi_+^{-2}(T)=m^2(T)=\frac{\l}{4}T_C^2(1+T/T_C)^2
\left(\frac{T-T_C}{T+T_C}\right),
\eeq
and
\beq
\label{eq:XI-}
\xi_-^{-2}(T)=-2m^2(T)=\frac{\l}{2}T_C^2(1+T/T_C)^2
\left(\frac{\left|T-T_C\right|}{T+T_C}\right).
\eeq
This is the well known result from mean-field theory, usually expressed as
\beq
\label{eq:XIMF}
\xi_{{\rm MF}}(T)\propto |T-T_C|^{-\nu}; \qquad \nu=1/2,
\eeq
where the critical exponent $\nu$ expresses the singular behavior of $\xi (T)$
as $T\rightarrow T_C$ both from above {\rm and} below. In Fig.~3 we show
the results for a dynamical simulation of a $d=2$ G--L model for various
values of $T$.\re{AAG}\ \ It is clear from this simulation that fluctuations
around the equilibrium value of $\f$ are indeed very large near $T_C$,
being considerably larger than the mean field results. Thus, the assumption of
near homogeneity is not valid if $T$ is sufficiently close to $T_C$.

In order to handle the infrared divergences that appear near $T_C$,
the RG is used to relate a given theory to an equivalent theory with
larger masses and thus better behaved in the infrared. Within the
$\ep$ expansion, one works in $4-\ep$ dimensions and finds a fixed
point of order $\ep$ of the RG equations, taking the limit
$\ep\rightarrow 1$ in the end. We refer the interested reader to
Ref.~\ref{BREZIN} for details.  To second-order in $\ep$ one obtains,
\beq
\label{eq:EPS}
\nu={1\over 2} + {1\over {12}}\ep + {7\over {162}}\ep^2 \simeq 0.63.
\eeq
The corrected critical exponent embodies corrections coming from the
infrared divergences near $T_C$. The $\ep$-corrected correlation length
can be written above $T_C$ as
\beq
\label{eq:XIE+}
\left [\xi_+^{\ep}(T)\right ]^{-1}=\sqrt{\frac{\l}{4}}\,
T_C\left (1+T/T_C\right )
\left( \frac{T-T_C}{T+T_C}\right)^{0.63}.
\eeq
Below $T_C$ we obtain,
\beq
\label{eq:XIE-}
\left [\xi_-^{\ep}(T)\right ]^{-1}=\sqrt{\frac{\l}{2}}\,
T_C\left (1+T/T_C\right )
\left( \frac{\left|T-T_C\right|}{T+T_C}\right)^{0.63},
\eeq
so that, in both cases the ratio between the mean field and $\ep$-corrected
correlation lengths can be written as
\beq
\label{eq:ETA}
{{\xi_{{\rm MF}}(T)}\over {\xi_{\ep}(T)}}=\e_C^{0.13}(T);\qquad
\e_C(T)\equiv {{|T-T_C|}\over {T+T_C}}.
\eeq

If we are interested in studying the behavior of the theory above $T_C$ we can
use the fact that $\xi(T)=m^{-1}(T)$ to obtain an $\ep$-corrected mass,
\beq
m_{\ep}(T)=\e^{0.13}_C(T)m(T) .
\eeq
A similar result can be easily obtained below $T_C$.

\vspace{48pt}
\thesection{\bf III. INFRARED CORRECTIONS TO THE ELECTROWEAK POTENTIAL}
\setcounter{section}{3}
\setcounter{equation}{0}
\vspace{18pt}

In this section we will argue that we can obtain information on the
critical behavior of the electroweak phase transition between $T_C$
and $T_2$ by studying a G--L model with a critical temperature that we
take to be $T_2$.  This is possible since $T_C$ is so close to $T_2$
due to the weakness of the transition already at 1-loop level. [See
Fig.~2.]  Thus, we will estimate the infrared corrections to the
electroweak model by looking at the G--L model around $T_C$.  Clearly
this is only an approximation to treating the full problem of
incorporating the $\ep$-expansion for the standard model.  However,
from the nature of the potential, we claim that our results are a
lower bound on the true infrared corrections, which we conjecture will
be even more severe than what we will estimate below.

\vspace{36pt}
\centerline{3.1 MATCHING TO A G--L MODEL ABOVE ITS CRITICAL TEMPERATURE}

We start with the simplest possible approach, by studying the G--L
model defined by the free energy density,
\beq
\label{eq:V1}
V_{{\rm G-L}}(\f,T)={{m^2(T)}\over 2}\f^2+
{{\l_T}\over 4}\f^4; \qquad m^2(T)\equiv 2D\left (
T^2-T_2^2\right ),
\eeq
where $D,~T_2$, and $\l_T$ are defined in the Introduction. This is
simply $V_{{\rm EW}}(\f,T)$ with $E\rightarrow 0$. This model exhibits
a second-order phase transition at $T=T_2$. Recall that this is the
temperature at which the barrier disappears in the 1-loop electroweak
potential. [See Fig.~1.] Thus, we are interested in the behavior of
this model for temperatures above $T_2$. The claim is that for $T\la
T_C$ and in the neighborhood of $\rf =0$ this model can be used to
give us an {\em estimate} of the infrared corrections to the
electroweak potential. Note that our choice of the mass is such that
the correlation length for fluctuations around equilibrium is the same
in both models.  In Fig.~4 we compare the electroweak potential and
the G--L model discussed above for $T=T_C$ and $T=T_2$.  Note how the
behavior around $\rf=0$ is well-matched by the G--L model

{}From the results of the previous section, the $\ep$-corrected mass is
\beq
\label{eq:MEP}
m_{\ep}^2(T)=2D\e_2^{0.26}(T)\left (T^2-T_2^2 \right ); \qquad
        \e_2(T)= {{|T-T_2|}\over {T+T_2}}.
\eeq
The value of $\eta_2(T)$ at $T=T_C$ can be found using $T_C$ and $T_2$
from Eqs.~(\ref{eq:TC}) and (\ref{eq:T2}):
\beq
\label{eq:ETAC}
\e_2(T_C)={{1-\sqrt{1-E^2/\l_TD}}\over {1+\sqrt{1-E^2/\l_TD}}}.
\eeq
In Fig.~5 we show $m_{\ep}^2(T_C)/m^2(T_C)=\eta_2^{0.26}(T_C)$ as a
function of the Higgs mass for several values of the top mass. It is
clear that the infrared corrections are quite large for all values of
parameters probed.  Below $T_C$ the potential is even flatter near the
origin and the infrared problem is even more severe. For larger values
of $\f$ the cubic term becomes important increasing the flatness of
the electroweak model compared to the G--L model (leading again to
more severe infrared problems). Before we go on to discuss possible
implications of the $\ep$-corrections to the electroweak phase
transition we study the same problem with a different G--L model next.
\vfill\eject
\vspace{36pt}
\centerline{3.2 MATCHING WITH G--L IN THE PRESENCE OF EXTERNAL FIELD}

Information about the critical behavior of the electroweak transition
by studying a simpler system can be obtained introducing a new field
such that the electroweak potential (neglecting the left-right
symmetry!) becomes equivalent to a G--L model with an external field
which is temperature dependent. In other words, we can transform away
the cubic term in $V_{{\rm EW}}(\f,T)$ by defining
\beq
\label{eq:FP}
\fp=\f-ET/\l_T,
\eeq
such that

\beq
\label{eq:VP}
V^{\prime}(\fp,T)=AT \left (T^2-T_C^2\right )\fp +B\left (T^2-T_C^{\prime 2}
\right )\f^{\prime 2} +{{\lambda_T}\over 4}\f^{\prime 4}+ \cdots
\eeq
where $\cdots$ represents terms independent of $\f^\prime$ and

\beq
A={{2ED}\over {\l_T}}\left (1-E^2/\l_TD\right ); \qquad B=D\left (1-3E^2/\l_TD
\right ).
\eeq
The temperature $T_C^\prime$ appearing in Eq.\ (\ref{eq:VP}) is given by
\beq
T_C^{\prime 2}={{T_2^2}\over {1-3E^2/2\l_TD}}\sim T_2^2(1+3x/2).
\eeq

The potential $V^{\prime}(\fp,T)$ is shown in Fig.~6 for several
values of the temperature. The coefficient of the linear term can be
interpreted as a temperature-dependent external field which vanishes
at $T_C$. At $T_C$ the potential has a double-well shape, so that
below $T_C$ the minimum at $\fp_-$ becomes metastable, while the
barrier separating $\fp_-$ from the global minimum at $\fp_+$
disappears at $T_2$.  Also, one can see that the location of the
minimum at $\fp_-$ is roughly temperature independent.  Hence, the
system exhibits the same critical behavior around $\fp_-$ as the
electroweak model around $\f =0$. We can see this by evaluating
\beq
m_-^2(T)={{\partial^2 V^{\prime}(\rf=\fp_-,T)}\over {\partial \f^{\prime 2}}}
=2B\left (T^2-T_C^{\prime 2}\right )+3\l_T\f_-^{\prime 2}.
\eeq
Using that at $T_C$ the potential is a double-well and that $\fp_-$ is roughly
temperature independent, we find that
\beq
\fp_-\simeq -\left [{{2B\left (T_C^{\prime 2}-T_C^2\right )}\over {\l_T}}
\right ]^{1/2}.
\eeq
Since $E^2/\l_TD \ll 1$ for all values of the Higgs and top masses we
consider, we can write
\beq
m_-^2(T)\simeq 2D\left (T^2-T_2^2\right )\left (1-{3\over 2}E^2/\l_TD\right )
\simeq m^2(T),
\eeq
where $m^2(T)$ is defined in Eq.~(\ref{eq:V1}).  Since we have shown
that $m_-^2(T)\simeq m^2(T)$, it is legitimate, within our framework,
to use the results from the $\ep$-expansion for the G--L model of $\fp$
directly into the electroweak model for $T\la T_C$ and in the
neighborhood of $\f=0$. As the relevant critical temperature of this
G--L model is also $T_2$, the results are identical to those of
Eq.~(\ref{eq:MEP}).

Again we stress that this is not intended to be an exact calculation
of the infrared corrections to the electroweak potential, but simply
an estimate of the magnitude of these corrections for small $\f$.  As
mentioned earlier, we expect the true corrections to be even more
severe that what we obtained above.

As a possible application of the above results, we estimate the
corrections to the 1-loop tunneling rate using $m_{\ep}^2(T)$. This is
clearly an approximation since we have stressed that our approach is
only valid in a small neighborhood of $\rf=0$, and should not be
trusted for $\f\ga D(T^2-T_2^2)/ET$, for a given $T$. We want to
estimate how severe the corrections to tunneling could be due to the
smallness of the curvature at the origin. The finite-temperature
tunneling rate, $\Gamma\propto\exp(-S_3/T)$, for a theory with a
potential like the electroweak potential has been shown by Dine {\it
et al.}\re{DINE} to have an approximate analytical expression for the
exponent given by
\beq
\frac{S_3}{T}=4.85\frac{m^3(T)}{E^2T^3}f(\alpha)\qquad
\alpha=\frac{\lambda_Tm^2(T)}{2E^2T^2},
\ee
with
\beq
f(\a)=1+{{\a}\over 4}\left [1+{{2.4}\over
{1-\a}}+{{0.26}\over {\left (1-\a\right )^2}}\right ].
\eeq
However, according to our arguments, for $T<T_C$ the effective
curvature of the potential around the equilibrium point is smaller
than what is estimated from the 1-loop approximation. The effective
tunneling barrier is then also smaller, and the kinetics of the
transition may be different from the usual nucleation scenario. [An
interesting possibility is that the critical temperature for the
corrected theory is larger than the 1-loop result. This is also true
for the results of Refs. \ref{BH} and \ref{DINE}, where the cubic term
is weakened due to the infrared corrections coming from the gauge
bosons. However, as we remarked earlier, our method is only applicable
in a small neighborhood of $\f$, and we cannot use it to study the
potential away from the origin which is necessary to predict $T_C$ in
a first order phase transition. We are presently investigating this
question.]  Taking into account the $\ep$-corrections above, the
exponent becomes,
\beq
\label{eq:RATE}
{{S_3^{\ep}}\over T}=
4.85\e_2^{0.39}(T)~{{m^3(T)}\over {E^2T^3}}f\left (\a_{\ep}\right );
 \qquad \a_{\ep}=\e^{0.26}_2(T)~{{\l_Tm^2(T)}\over {2E^2T^2}}.
\eeq

Clearly use of the $\ep$-expansion improved mass can have an enormous
effect upon the tunnelling rate, changing the {\em exponent} by a
large factor.

\vspace{48pt}
\thesection{\bf IV. THERMAL FLUCTUATIONS AND SUB-CRITICAL BUBBLES}
\setcounter{section}{4}
\setcounter{equation}{0}
\vspace{18pt}

In this section we discuss how it is possible to examine the strength
of a first-order transition by two simple methods. The first method,
introduced by one of us,\re{GLEISER} relies on estimating the
magnitude of the thermal dispersion of the order parameter around its
equilibrium value. The second method based on the work of Gleiser,
Kolb and Watkins,\re{GKW} relies on estimating the thermal nucleation
rate of ``sub-critical bubbles,'' which are correlation volume
fluctuations of one phase inside the other phase.  We will argue that
when applied to the electroweak phase transition, both methods give
results which are qualitatively consistent with each other, signaling
the failure of the na\"{\i}ve 1-loop potential as a valid
approximation to study the dynamics of the transition.

\vspace{36pt}

\centerline{4.1 THERMAL DISPERSION AROUND EQUILIBRIUM}

Consider a system described by some potential which at a given
temperature $T$ exhibits at least a local minimum at some value of the
local order parameter, $\f=\rf$. For example, the electroweak
potential of Fig.~1 has a global (and, below $T_C$ local) minimum at
$\rf=0$ for all $T> T_2$. If we are interested in studying nucleation
below $T_C$, we should require that the system is in a near
homogeneous phase characterized by $\rf=0$. That is, although there
will be fluctuations around equilibrium, they should be small enough
so that when calculating the transition rate the usual boundary
conditions at infinity apply.  However, from our previous discussion,
for weakly first-order transitions infrared corrections can be
important, and large fluctuations around equilibrium are to be
expected for temperatures below $T_C$.  It is then legitimate to ask
if the usual assumption of near-homogeneity is valid.  In Ref.\
\ref{GLEISER} a simple method was introduced in order to answer
qualitatively this question. Using a Gaussian approximation to the
potential, we compare the thermal dispersion of the order parameter
around its equilibrium value for a certain length scale $r=|x-y|$,
denoted by $\langle \f(x)\f(y)\rangle_{\beta}$, to the value of the
order parameter at the inflection point of the potential, denoted by
$\f_{{\rm inf}}$. Simple physical arguments show that the relevant
length scale is the correlation length, $\xi$.\re{GLEISER}\ \ If the
magnitude of the thermal dispersion is comparable to the value of $\f$
at the inflection point, the system has a large probability to
overcome the barrier thermally, populating other accessible minima. In
equations, if we have
\beq
\label{eq:GINZ}
\sqrt{ \langle \f(0)\f(\xi)\rangle}_{\beta}\la |\f_{{\rm inf}} - \rf |,
\eeq
the assumption of near-homogeneity is probably incorrect. In Ref.\
\ref{GLEISER} this condition was applied to the top of the barrier
instead of the inflection point. That gives a very conservative
estimate, since at the top of the barrier non-linearities are
obviously important, enhancing the magnitude of the thermal
dispersion. Strictly speaking, the Gaussian approximation is only
valid up to the inflection point.

In a recent work, Tetradis applied the above method to the electroweak
transition showing that indeed the probability of fluctuations over
the barrier is large.\re{TETRADIS}\ \ He used an approximate
expression for calculating $\langle \f^2\rangle_{\beta}$ given in
Ref.\ \ref{GLEISER}. Here we would like to compute the thermal
dispersion in more detail, and again apply the results to the
electroweak transition. We will show that the dispersion is still
quite large, in qualitative agreement with our previous results based
on the $\ep$-expansion.

For a free massive scalar field theory in thermodynamic equilibrium at
temperature $T$, the two-point function can be written in terms of a
zero temperature part and a finite temperature part after time
ordering as\re{IZ}
\ba
\label{eq:2POINT}
\langle {\widehat T}\f(x) \f(y) \rangle & = & i\int {{d^4\!k}\over {(2\pi)^4}}
{{e^{-ik(x-y)}}\over {k^2-m^2+i\ep}}\nonumber\\
& & +\int {{d^3\!k}\over {(2\pi)^32\omega_k}}
{1\over {e^{\beta\omega_k}-1}}\left (e^{-ik(x-y)}+e^{ik(x-y)}\right ),
\ea
where $\omega_k^2=\bk^2+m^2$.

Here we will focus on the temperature dependent part only. In spherical
coordinates we can write
\beq
\langle\f(0)\f(|\br|)\rangle_{\beta}\equiv\Delta(|\br|,\beta)=
\int {{d|\bk||\bk^2|}\over {(2\pi)^2\omega_k}}{{{\rm cos}\left (
|\bk||\br|{\rm cos}\theta \right ){\rm sin}\theta d\theta}\over
{e^{\beta\omega_k}-1}}.
\eeq
Using properties of Bessel functions $\Delta(\br,\beta)$ can be written as
[for details see Ref.\ (\ref{KW})],
\beq
\label{eq:DELTA}
\Delta(|\br|,\beta)={m\over {2\pi^2}}\sum_{n=1}^{\infty}{{{\rm K}_1\left [
m\beta \sqrt{n^2+\left (|\br|/\beta\right )^2}\right ]}\over
{\sqrt {n^2+\left (|\br|/\beta\right )^2}}},
\eeq
where ${\rm K}_1[z]$ is the modified Bessel function of first kind.
{}From the asymptotic properties of Bessel functions one obtains, in the
high temperature limit,
\beq
\Delta(0,m\beta \ll 1)\simeq {{T^2}\over {12}},
\eeq
where we used that $\beta =T^{-1}$. We are interested in fluctuations
of a correlation volume at a temperature $T$. In the spirit of the
Gaussian approximation we will take the mass $m$ in the above formula
to be the curvature of the potential around its equilibrium point at a
temperature $T$, as defined in Eq.\ (\ref{eq:XI}).  [This is precisely
what is done in the traditional analysis of fluctuations in a
Ginzburg--Landau model.\re{LANDAU}] The correlation length is then
simply $\xi=m^{-1}$ and we can write
\beq
\Delta(\xi(T),T)={{T^2}\over {2\pi^2}}\sum_{n=1}^{\infty}{x\over
{\sqrt {n^2+x^{-2}}}}{\rm K}_1\left [x\sqrt {n^2+x^{-2}}\right ] ;
\qquad x\equiv m/T.
\eeq

The probability that a fluctuation of correlation volume around
equilibrium can ``spread'' over the inflection point is then simply,
\beq
\label{eq:PROB}
{\rm P}\left (\rf\rightarrow \f_{{\rm inf}}\right )\sim
{\rm exp}\left [-{{\left (\rf-\f_{{\rm inf}}\right )^2}\over
{2\Delta(\xi(T),T)}}\right ],
\eeq
where it should be clear that $\rf$ and $\f_{{\rm inf}}$ are in
general temperature dependent quantities.  It is straightforward to
apply this formula to the electroweak transition at $T_C$. $m^2(T)$ is
given in Eq.~(\ref{eq:V1}) and $\f_{{\rm inf}}$ for $T\leq T_1$ is
given by
\beq
\f_{{\rm inf}}={{ET}\over {\l_T}}\left [1-\sqrt {1-{{2D\l_T}\over {3E^2}}
\left (1-T_2^2/T^2\right ) }\right ]~~.
\eeq
The results are shown in Fig.~7 as a function of the Higgs mass for
$M_T=130$ GeV. It is clear that thermal fluctuations are quite large
at $T_C$, in agreement with our previous results. Since the 1-loop
potential is evaluated for small fluctuations around the equilibrium
value, this simple criterion indicates that the 1-loop approximation
to the effective potential is not reliable around $T_C$.

\vspace{36pt}

\centerline{4.2 SUB-CRITICAL BUBBLES}

We now discuss another criterion to estimate the validity of the 1-loop
approximation to the effective potential, based on the ``sub-critical
bubbles method'' discussed in Ref.~\ref{GKW}. Since we have recently
applied this method to the electroweak potential of Eq.~(\ref{eq:VEW}), we
will be quite brief here and refer the reader to Ref.\ \ref{GK} for details.

Consider the electroweak potential of Fig. 1. Below $T_1$ a new
minimum develops at $\f_+$ away from the symmetric minimum at $\rf=0$.
There will be a non-zero probability for bubbles of radius $R$ of the
new phase at $\f_+$ to be thermally nucleated. The thermal nucleation
rate for producing a bubble of radius $R$ is given by $\Gamma(R,T)\sim
{\rm exp}[-F(R)/T]$, where $F(R)$ is the free energy of the
fluctuating region of radius $R$. For $T\geq T_C$ it is clear that the
larger the bubble the more unfavored it is, since the free energy is a
monotonically increasing function of $R$. The bubbles will shrink in a
time scale determined by many factors. For example, for a curvature
dominated motion of the bubble wall, which is probably a good
approximation close to $T_C$, the radius of a large bubble shrinks as
$t^{1/3}$.\re{FISHER} Some recent numerical studies showed that even
small bubbles persist longer than one would na\"{\i}vely estimate,
bouncing back a few times before dissipating all their energy into
quanta of the field.\re{SRIVASTAVA} However, due to the exponential
suppression in their production rate, unless the transition is very
weakly first order (with the whole bubble picture being invalid in
this case), only bubbles with small enough radius can be efficiently
produced so that at any given time a reasonable fraction of the
Horizon volume can be occupied by the new phase at $\f_+$. Although
there is a distribution of bubbles with different radii, it is clear
from the above arguments that bubbles with a correlation volume will
be statistically dominant. [The kinetics of the transition is bound to
be much more complicated than these simple arguments may imply. There
will be many different processes contributing to the number density of
bubbles of a given radius as a function of time, such as capture and
evaporation of particles from bubbles, coalescence due to bubble
collisions, shrinking of larger bubbles, neighbor-induced nucleation,
and possible shape instabilities, to name just a few.\re{GG} A quick
glance at Fig. 3 should convince the reader of this.]

The basic idea behind the sub-critical bubbles method is that for
sufficiently weak first order transitions, the rate for producing
bubbles of a correlation volume is quite large, so that at any given
time there will be an appreciable fraction of the total volume
occupied by the new phase. If this is the case, the usual assumption
of near-homogeneity used in vacuum decay calculations is not valid;
instead of having critical bubbles being nucleated on a background of
the metastable phase, nucleation would occur in a background which is
better described by a dilute gas of small, non-perturbative
fluctuations.  There is no reason to expect that the usual calculation
for the decay rate is applicable in this case.  This method
complements the estimates for the thermal dispersion around
equilibrium discussed in Section 4.1, with the important difference
that the sub-critical bubble calculations do include the
non-linearities in the problem, being by definition non-perturbative.

The free energy of a spherically symmetric fluctuation around equilibrium
is
\beq
F(T)=4\pi\int_0^{\infty}r^2~dr\left [{1\over 2}\left ({{d\f}\over {dr}}
\right )^2 + V_{{\rm EW}}(\f,T)\right ] .
\eeq
We will focus on the electroweak model at $T_C$. In principle, there
will be fluctuations from $\f=0$ to $\f_+$ and back, although at $T_C$
the free energies for these fluctuations are identical. The rates for
the thermal fluctuations can be estimated by making an ansatz for the
radial profile of the sub-critical bubbles. Following Ref.\ \ref{GK}
we write
\beq
\f_+(r)=\f_+{\rm exp}\left (-r^2/\ell^2\right ),
\eeq
vwhere $\f_+(r)$ corresponds to a bubble of broken phase $\f_+$ nucleated
in the symmetric phase $\f=0$. The parameter $\ell$ controls the approximate
size of the bubble which we take to be the correlation length. Introducing the
dimensionless variables $X(\rho)=\f(r)/\sigma,~\tilde \ell(T)=\ell(T)\sigma,~
\theta=T/\sigma,$ and $\rho=r\sigma$, we obtain
\beq
F_+(\theta)=\pi^{3/2}X_+^2\tilde\ell\sigma\left [{{3\sqrt{2}}\over 8}+
\tilde\ell^2\left ({{D\sqrt{2}}\over 4}\left (\theta_C^2-\theta_2^2\right )-
{{E\theta_C\sqrt{3}}\over 9}X_+ + {{\l_T}\over {32}}X_+^2\right )\right ].
\eeq
In Fig. 8 we show $F_+(T_C)/T_C$ as a function of the Higgs mass and
the top mass. In order for sub-critical bubbles to be of cosmological
relevance, their thermal nucleation rate must be considerably larger
than the expansion rate of the Universe, $\Gamma(\xi(T),T)/H> 1$, with
$H\simeq 1.66g_{\ast}^{1/2}T^2/M_{{\rm PL}}$, where $g_{\ast}\simeq
110$ is the number of effective relativistic degrees of freedom at the
electroweak scale. Neglecting pre-factors, this condition can be
easily seen to lead to the inequality $F_+(T)/T<34$. From Fig. 8 it is
clear that at $T_C$ this condition is comfortably satisfied for the
present lower bound on the Higgs mass, $M_H\geq 57$ GeV, for which we
obtain $\Gamma(\xi,T_C)/H\sim 10^{8}$.

Recently, Dine {\it et al.} argued that sub-critical bubbles would
not be of relevance for most (if not all) the parameter space of the
standard model due to the smallness of the thermal dispersions around
$\rf=0$.\re{DINE} We agree with their results for $M_H\sim 60$ GeV.
However, for larger Higgs masses fluctuations in long wavelengths are
quite large, contrary to their claim.  We hoped to have shown here
that both the estimate from the thermal dispersion and from
sub-critical bubbles indicate that there will be large fluctuations
around equilibrium, signaling the failure of the 1-loop potential to
describe the dynamics of the transition.

\vspace{48pt}
\thesection{\bf V. CONCLUSION}
\setcounter{section}{5}
\setcounter{equation}{0}
\vspace{24pt}

In this work we have argued that it is possible to study the critical
behavior of a weak first order transition which has a spinodal
instability at some temperature $T_2$ by mapping its behavior around
equilibrium, $\rf$, to an effective Ginzburg-Landau model above its
critical temperature $T_2$. In this way, both models have the same
spinodal instability at $\rf$ so that infrared corrections can be
estimated from well-known $\ep$-expansion methods. This approach is
completely general and can in principle be applied to any sufficiently
weak first order transition. It suits the standard electroweak model
particularly well due to the closeness of its critical temperature
$T_C$ to the spinodal instability temperature $T_2$. In fact, the
difference between the two temperatures should provide a qualitative
measure of the weakness of the transition.

Incorporating the $\ep$-expansion results leads to a larger
correlation in the spatial fluctuations of the order parameter, which
can be translated into a smaller (infrared corrected) mass for
excitations around $\rf$.  Thus, the strength of the transition is
considerably weaker than one would estimate from the na\"{\i}ve 1-loop
potential. We do not claim here to have obtained the $\ep$-corrected
effective potential, but an estimate of the infrared corrections which
are not included in the 1-loop result. Our results provide a simple
way to examine the importance of these corrections around $T_C$,
offering a simple way of estimating the strength of the transition. If
the $\eta$ parameter is close to unity at the critical temperature
$T_C$ the transition is well described by the 1-loop result.
Otherwise, the transition is weakly first order, and one should be
very careful when adopting the usual vacuum decay formalism to study
the transition. As a pictorial representation of the complexity of the
behavior of a weakly first order transition, we show in Fig.\ 3 the
behavior of a $\f^4$ model around $T_C$. For clarity, we show only the
black and white values of $\f$, defined to be for $\f<0$ and $\f\geq
0$, respectively. The critical behavior of the system is then in the
same equivalence class as the 2-dimensional Ising model.  This
simulation is done by starting the system at $\f=+\f_0$ and $T=0$, and
by studying the temperature behavior of the system by immersing it in
a thermal bath at temperature $T$.\re{AAG} For temperatures just below
or above $T_C$, the large fluctuations around equilibrium are quite
apparent. In particular, one can picture the behavior of this system
just below its critical temperature as being qualitatively similar to
the behavior of a weak first-order transition {\it at} its critical
temperature.

We have also discussed two other simple ways of estimating the
strength of the transition based on the thermal dispersion around
equilibrium and on the sub-critical bubbles method. When applied to
the 1-loop electroweak potential both approaches suggest that there
will be large fluctuations around equilibrium, indicating that the
1-loop result does not fully describe the dynamics of the transition.
In fact, the results here show that the actual dynamics of the
transition may be much more complex than the usual scenario based on
vacuum decay calculations.

\vspace{36pt}

\centerline{ACKNOWLEDGEMENTS}

We would like to thank F.\ Alcarraz, R.\ Holman, J.\ Langer, and S.\
Shenker for important discussions. We also thank the hospitality of
the Institute for Theoretical Physics at Santa Barbara where this work
was developed during the Cosmological Phase Transitions workshop. This
work was partly supported in Santa Barbara by the National Science
Foundation under Grant No. PHY89--04035.  EWK was partly supported by
NASA under Grant NAGW--2381 at Fermilab and by the DOE at Chicago and
Fermilab.

\vspace{1.0in}
\centerline{{\bf References}}
\frenchspacing
\begin{enumerate}

\item\label{KRS} V. A. Kuzmin, V. A. Rubakov,
and M. E. Shaposhnikov, {\em Phys. Lett.} {\bf 155B}, 36 (1985).

\item\label{EWB} M. E. Shaposhnikov, {\em Nucl. Phys.} {\bf B287},
757 (1987); P. Arnold and L.  McLerran, {\em Phys. Rev. D} {\bf 36},
581 (1987); M. Dine, O.  Lechtenfeld, B. Sakita, W. Fischler, and J.
Polchinski, {\em Nucl.  Phys.} {\bf B342}, 381 (1990); N. Turok and J.
Zadrozny, {\em Phys.  Rev. Lett.} {\bf 65}, 2331 (1990); {\em Nucl.
Phys.} {\bf B358}, 471 (1991); A. G. Cohen, D. B. Kaplan, and A. E.
Nelson, {\it Phys. Lett.} {\bf 245B} (1990) 561; {\em Nucl. Phys.}
{\bf B349},727 (1991); M.  Dine, P. Huet, R. S. Singleton Jr., and L.
Susskind, {\em Phys. Lett.} {\bf 257B}, 351 (1991).

\item\label{CKN} A. G. Cohen, D. B. Kaplan, and A. E. Nelson, in Ref. 2.

\item\label{TZ} N. Turok and J. Zadrozny, in Ref. 2.

\item\label{FINITET} L. Dolan and R. Jackiw, {\em Phys. Rev. D} {\bf
9}, 3320 (1974); S. Weinberg, {\it ibid.}, 3357 (1974).

\item\label{AH} G. W. Anderson and L. J. Hall, {\em Phys. Rev. D} {\bf 45},
2685 (1992); M. Dine, P. Huet, and R. Singleton, {\em Nucl. Phys.}
{\bf B375}, 625 (1992).

\item\label{BH} D. Brahm and S. D. H. Hsu, Caltech preprints, CALT-68-1705
and CALT-68-1762 (1991); M. E. Shaposhnikov, {\em Phys. Lett.} {\bf B277},
324 (1992); M. E. Carrington, {\em Phys. Rev. D} {\bf 45}, 2933 (1992).

\item\label{DINE} M. Dine, R. Leigh, P. Huet, A. Linde, and D. Linde,
{\em Phys. Rev. D} {\bf 46}, 550 (1992).

\item\label{ARNOLD} P. Arnold, University of Washington preprint UW/PT-92-06
(1992).

\item\label{DEGENNES} P. G. De Gennes, {\it The Physics of Liquid Crystals},
(Clarendon, Oxford, [First edition, reprinted 1975]).

\item\label{POTTS} L. A. Fern\'andez, J. J. Ruiz-Lorenzo, M. Lombardo, and
A. Taranc\'on, {\em Phys. Lett.} {\bf B277}, 485 (1992).

\item\label{WILSON} See, for example, K. G. Wilson and J. Kogut, {\em Phys.
Rep.} {\bf 12C}, 75 (1974).

\item\label{GINSPARG} P. Ginsparg, {\em Nucl. Phys.} {\bf B170}[FS1],
388 (1980).

\item\label{BREZIN} E. Br\'ezin, in {\it Methods in Field Theory}, Les
Houches 1975, Session XXVIII, ed. R. Balian and J. Zinn-Justin, (North-Holland,
Amsterdam 1976).

\item\label{AAG} F. F. Abraham, M. G. Alford, and M. Gleiser,

{\it in progress}.

\item\label{GLEISER} M. Gleiser, {\em Phys. Rev. D} {\bf 42}, 3350 (1990).

\item\label{GKW} M. Gleiser, E. W. Kolb, and R. Watkins, {\em Nucl. Phys.}
{\bf B364}, 411 (1991).

\item\label{TETRADIS} N. Tetradis, DESY preprint DESY 91-151 (1991).

\item\label{IZ} C. Itzykson and C. B. Zuber, {\it Quantum Field Theory},
(McGraw-Hill, New York, 1980).

\item\label{KW} E. W. Kolb and Y. Wang, {\em Phys. Rev. D} {\bf 45},
4421 (1992).

\item\label{LANDAU} L. D. Landau and E. M. Lifshitz, {\it Statistical Physics},
3rd ed. (Pergamon, New York, 1980), Pt. I.

\item\label{GK} M. Gleiser and E. W. Kolb, Fermilab preprint

FERMILAB-PUB-91/305-A (1991), in press {\it Phys. Rev. Lett.}

\item\label{FISHER} D. A. Huse and D. S. Fisher, {\em Phys. Rev.} {\bf B35},
6841 (1987); C. Tang, H. Nakanishi, and J. S. Langer, {\em Phys. Rev.}
{\bf A40}, 995 (1989).

\item\label{SRIVASTAVA} A. M. Srivastava, University of Minnesota preprint,
TPI-MINN-91/37-T (1991).

\item\label{GG} G. B. Gelmini and M. Gleiser, {\it in progress}.

\end{enumerate}

\newpage
\centerline{\bf Figure Captions}

FIG.~1. The 1-loop electroweak potential at several different temperatures.

FIG.~2. The parameter $x=E^2/\l_TD$ as a function of the Higgs mass for
several values of the top quark mass.

FIG.~3. 2-dimensional simulation of $\f^4$ model for different temperatures.
Initial conditions were chosen such that at $t=0$ the field is in the
equilibrium state $\f=+\f_0$, given by the G--L model at $T=0$.

FIG.~4. The 1-loop electroweak potential (solid curves) and the
associated G--L model of Section 3.1 (dashed curves) for temperatures
$T_C$ (a) and $T_2$ (b). A Higgs mass of 100 GeV and a top quark
mass of 130 GeV were chosen.

FIG.~5. The $\ep$-corrected mass as a function of the Higgs mass for several
values of the top mass.

FIG.~6. The G--L model obtained by neglecting the left-right symmetry in the
electroweak model and by transforming away its cubic term.

FIG.~7. The ratio $\f_{{\rm inf}}^2/2\Delta(\xi )$  as a function of the
Higgs mass for $M_T=130$ GeV at $T_C$.

FIG.~8. The free energy of the sub-critical fluctuation at the critical
temperature as a function of the Higgs mass for several values of the top quark
mass.

\end{document}